# Fermi Surface and Light Quasi Particles in Hourglass Nodal Chain Metal β-ReO$_2$

Daigorou Hirai[1,2], Takahito Anbai[2], Takako Konoike[3], Shinya Uji[3], Yuya Hattori[3], Taichi Terashima[3], Hajime Ishikawa[2], Koichi Kindo[2], Naoyuki Katayama[1], Tamio Oguchi[4,5], and Zenji Hiroi[2]

[1] Department of Applied Physics, Nagoya University, Nagoya 464–8603, Japan
[2] Institute for Solid State Physics, University of Tokyo, Kashiwa, Chiba 277-8581, Japan
[3] International Center for Materials Nanoarchitectonics, National Institute for Materials Science, Tsukuba, Ibaraki 305-0003, Japan
[4] Center for Spintronics Research Network, Graduate School of Engineering Science, Osaka University, Toyonaka, Osaka 560-8531, Japan
[5] Spintronics Research Network Division, Institute for Open and Transdisciplinary Research Initiatives, Osaka University, Suita, Osaka 565-0871, Japan

E-mail: dhirai@nuap.nagoya-u.ac.jp

## Abstract

Quantum oscillations in magnetic torque and electrical resistivity were measured to investigate the electronic structure of β-ReO$_2$, a candidate hourglass nodal chain metal (Dirac loop chain metal). All the de Haas-van Alphen oscillation branches measured at 30 mK in magnetic fields of up to 17.5 T were consistent with first-principles calculations predicting four Fermi surfaces (FSs). The small-electron FS of the four FSs exhibited a very small cyclotron mass, 0.059 times that of the free electrons, which is likely to be related to the linear dispersion of the energy band. The consistency between the quantum oscillation results and band calculations indicates the presence of the hourglass nodal chain predicted for β-ReO$_2$ in the vicinity of the Fermi energy.

## 1. Introduction

In recent years, considerable attention has been paid to topological semimetals that exhibit remarkable transport properties, responses to magnetic fields, and surface states[1]–[5]. Topological semimetals are characterised by band crossings near the Fermi energy ($E_F$); Weyl and Dirac semimetals have doubly and quadruply degenerate band-crossing points, where the quasiparticles are described as Dirac and Weyl fermions.

In nodal line semimetals, the band crossing is not a point in $k$-space but a one-dimensional loop—the so-called nodal line. In particular, nodal lines produce drumhead surface states[6]. This surface state has a high density of states, and surface superconductivity and ferromagnetism may manifest accordingly[7],[8].

Generally, spin-orbit interactions (SOIs) open an energy gap at the band crossings; however, in topological semimetals, band crossings are protected from the gap opening owing to the symmetry of the crystal structure[9]. For example, in the representative Dirac semimetals Na$_3$Bi[10],[11] and Cd$_3$As$_2$[12]–[14], the Dirac point along the rotation axis is protected by the rotation symmetry. In the nodal-line semimetals Ca$_3$P$_2$[15],[16] and CaAgX (X = P, As) [17], the nodal lines are protected within the mirror plane owing to the reflection/mirror symmetry.

Reportedly, materials have band crossings that are protected by non-symmorphic symmetries including partial translations such as glide planes and screw axes, rather than symmorphic symmetries like rotation and mirror symmetry. In materials with glide symmetry, bands may exchange pairs within the glide plane, thereby forming band dispersions with an hourglass shape; the Dirac point at the neck-crossing point of the hourglass is protected against SOIs through glide symmetry[18],[19]. Such hourglass dispersions have been observed in KHgSb using angle-resolved photoemission spectroscopy (ARPES)[20]. Nodal lines protected by glide symmetry have also been suggested for SrIrO$_3$[21], WHM (W = Zr, Hf, La; H = Si, Ge, Sn, Sb; M = O, S, Se, Te)[22], and IrO$_2$[23].

In materials with multiple orthogonal glide planes, multiple nodal lines (Dirac loops) appear that are protected by each glide plane, which may be connected by a single point, forming a chain-like structure in *k*-space. Materials with such electronic structures are called nodal-chain (NC) metals: these materials represent a new category of topological semimetals[24]. Wang et al. have reported that in a material with an α-PbO$_2$-type crystal structure (space group *Pbcn*), two orthogonal nodal lines protected by *n*- and *b*-glide are connected at a single point and possess an NC along the $k_y$ direction[25]. Notably, in β-ReO$_2$, the first-principles calculations have predicted NCs near the $E_F$: the NC metal is called the "Dirac loop chain metal". Thus, unusual transport properties and surface states derived from NCs are expected for β-ReO$_2$.

Particularly, only the metallic electrical conduction in β-ReO$_2$ has been reported[26]; its magnetotransport and other properties are yet to be characterised. In addition, single crystals used in the previous study exhibited a fairly high residual resistivity of 10 μΩcm. To investigate the NC signature in β-ReO$_2$, we grew high quality single crystals with an extremely low residual resistivity of 206 nΩcm and measured their physical properties[27]. A large transverse magnetoresistance of 22,000% was observed at 2 K in a field of 10 T. In addition, quantum oscillations (QOs) were observed at high temperatures and low magnetic fields of 7 K and 7 T, indicating the existence of quasiparticles with cyclotron masses that were 0.4 times lighter than free electrons. A large magnetoresistance and the presence of light quasiparticles are common features of topological semimetals. Therefore, the presence of Dirac electrons in β-ReO$_2$ is strongly suggested. However, it is not clear whether the large magnetoresistance and light quasiparticles are derived from NCs, because there are several Fermi surfaces (FSs) other than those associated with NCs, as suggested by first-principles calculations[27]. To confirm the presence of NCs in β-ReO$_2$, the FSs must be experimentally measured, and the electronic structure must be determined.

In this study, QOs in the magnetic torque and electrical resistivity were measured and compared with first-principles calculations to investigate the FSs of β-ReO$_2$ in detail. The observed QO frequencies corresponded to the extremal cross-sectional areas of all four FSs predicted by first-principles calculations, and the angular dependences of the QO branches between the calculations and experiments were consistent. The cyclotoron mass of the quasiparticles estimated from the temperature dependence of the oscillation amplitude was approximately identical to that of free electrons for the three large FSs, whereas for the small electron pocket, the cyclotoron mass was extremely light—0.059 times lighter than free electrons. The shape of the electron pocket surrounded by one of the two nodal lines forming the NC was determined with high accuracy, and first-principles calculations revealed that the NC was located extremely close to the $E_F$ in β-ReO$_2$.

## 2. Experimental

Single crystals of β-ReO$_2$ were grown by chemical vapor transport method using iodine as a transport agent as in previous reports[27]. Obtained crystals larger than 1 × 0.3 × 0.3 mm$^3$ were mostly twinned, while some of those smaller than 0.3 × 0.1 × 0.1 mm$^3$ were single-domain crystals. Single crystal X-ray diffraction (XRD) experiments (R-AXIS RAPID II RIGAKU) were used to confirm the α-PbO$_2$-type crystal structure with space group *Pbcn*, select single-domain crystals, and determine their crystal orientation.

Magnetic torque measurements were performed by a piezo-micro-cantilever technique[28]. A single crystal was attached to the cantilever by silicon grease, and the torque signals were detected by a lock-in amplifier at a frequency of 15 Hz using a homemade bridge circuit. This technique allows for highly accurate measurements of small magnetic torque signals. The measurements were performed at the Tsukuba Magnet Laboratories of NIMS using a 20 T superconducting magnet and a dilution refrigerator.

Electrical resistivity measurements were performed using a 17 T superconducting magnet and a $^4$He gas-flow cryostat. Only single-domain portions cut from a large single crystal were used for resistivity measurements. The sample was confirmed to be a single domain by single-crystal XRD measurements.

The electrical resistivity up to approximately 56 T was measured in the pulsed high magnetic field by a four terminal AC method at the frequency of 20 kHz. The voltage was recorded by a digital ocilloscope at the sampling rate of 1 MS/s and analyzed by the numerial phase detection technique[29]. The pulsed magnetic field with the duration of 40 milliseconds was generated by a multi-layer coil at the International MegaGauss Science Laboratory at ISSP.

For structural analysis, powder XRD experiments were conducted at 150 K and X-ray energy of 20 keV using a quadruple PILATUS 100K detector at the BL5S2 of Aichi Synchrotron Radiation Center. The sample was prepared by crushing single crystals and sealing it in a Lindemann capillary of 0.1 mm diameter. RIETAN-FP was used for the Rietveld analysis[30].

First-principles electronic structure calculations were performed using the all-electron full-potential linearized augmented plane wave method[31]–[33] implemented in the HiLAPW code[34] with the Perdew-Burke-Ernzerhof generalized gradient approximation to the density functional theory[35]. The SOIs were self-consistently taken into account for the valence and core states by the second variation scheme[36]. The energy cutoffs of 20 and 160 Ry were used for wavefunction and potential expansions, respectively. The lattice constants and atomic coordinates were taken from experimental data obtained by a Rietveld refinement for the synchrotron powder X-ray diffraction pattern in this study, in contrast to the previous calculations[27] which used structural parameters optimised *via* first-principles calculations[37]). Brillouin-zone sampling was made by the tetrahedron integration scheme with $\Gamma$-centered $16 \times 16 \times 16$ ($32 \times 32 \times 32$) mesh points in the self-consistent-field (density of states) calculations. Fermi surfaces were drawn with "FermiSurfer" program[38]).

## 3. Results

### 3.1 dHvA oscillation in magnetic torque curve

Figure 1a shows the magnetic-field dependence of the magnetic torque of $\beta$-ReO$_2$ at 30 mK and $B \parallel$ [110]. The magnetic torque is given by $\tau_{torque} = \boldsymbol{M} \times \boldsymbol{B}$, where $\boldsymbol{M}$ and $\boldsymbol{B}$ are the magnetisation and magnetic field, respectively. The magnetic torque comprises a monotonically increasing and an oscillating components, the former originates from Pauli paramagnetic magnetisation, whereas the latter corresponds to the QO. The QO component (upper curve) is extracted from the raw data by subtracting the Pauli paramagnetic component that is expressed by a polynomial function. The oscillating components exhibit typical de Haas-van Alphen (dHvA) oscillation behaviour, with amplitudes increasing with the magnetic field. Upon enlarging the high-field region, the oscillations are observed to exhibit a saw-tooth shape. This is caused by mixing of the harmonic components of the fundamental oscillation in high-purity crystals. Many dHvA peaks up to a relatively high frequency of 3300 T are present in the fast Fourier transform (FFT) spectrum of the QO component (Fig. 1b). Each dHvA peak is labelled from α to ζ and their composite components, starting from the lowest magnetic field. As expected from the oscillatory waveform, harmonic components up to the third order of the frequency of δ are present herein.

Based on the Onsager relation, the observed frequency $F$ in the FFT spectrum yields the extremal cross-sectional area of the FS perpendicular to the magnetic field $A_F$: $F = (\Phi_0/2\pi^2)A_F$, where $\Phi_0$ is the magnetic quantum flux. The highest fundamental frequency labbeled as ζ is 3020 T, corresponding to the cross-sectional area of the FS $A_F = 0.288$ Å$^{-2}$. This area is considerbaly large; 49% of the cross-sectional area perpendicular to [110] in the first Brillouin zone, 0.586 Å$^{-2}$. However, the lowest frequency, α, is approximately 70 T, which originates from a cross-sectional area of less than 1/40th of ζ.

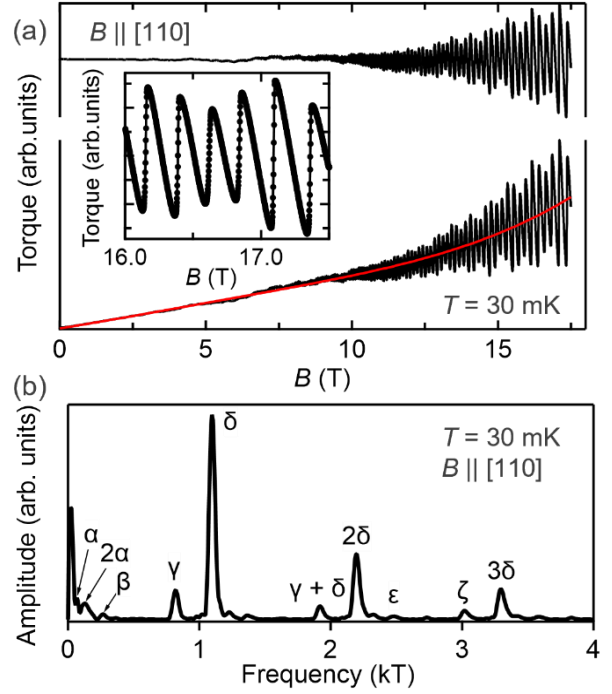

**Figure 1.** (a) Magnetic-field dependences of magnetic torque for a $\beta$-ReO$_2$ crystal at $T = 30$ mK and the magnetic field parallel to the [110] direction. The raw, background curve, and oscillatory component after subtracting background component are depicted by the black and red lines in the lower panel and the black line in the upper panel, respectively. The inset shows a magnified range at high fields to demonstrate quantum oscillation. (b) Fast Fourier transform spectrum of the dHvA oscillation in the field range between 5 and 10 T. The observed peaks are assigned as α, β, γ, δ, ε, ζ, and their higher order summations.

### 3.2 Angle dependence of dHvA branches

The magnetic-field-angle dependence of dHvA oscillations in the magnetic-field orientations rotated around the *c* and *b* axes are displayed in Fig. 2: the angles are defined as $\theta_1$ and $\theta_2$, $\theta_1 = \theta_2 = 0°$ for $B \parallel a$, $\theta_1 = 90°$ for $B \parallel b$, and $\theta_2 = 90°$ for $B \parallel c$, respectively. Figure 2 shows only the fundamental frequency, excluding the synthetic and harmonic components. $B \parallel$ [110] corresponds to $\theta_1 = 49.6°$, and the branches are labelled in Fig. 2 based on the assignment in Fig. 1. Most branches have a minimum at $B \parallel a$, and the frequency increases towards both $B \parallel b$ and $B \parallel c$. This suggests the presence of a cylindrical or ellipsoidal Fermi surface extending parallel to the *a* axis; for a spherical surface, the extremal cross-sectional area is constant for any angles. The highest frequencies of ζ assume maxima values of 3317 and 3555 T at $\theta_1 = 64°$ and $\theta_2 = 54°$, respectively.

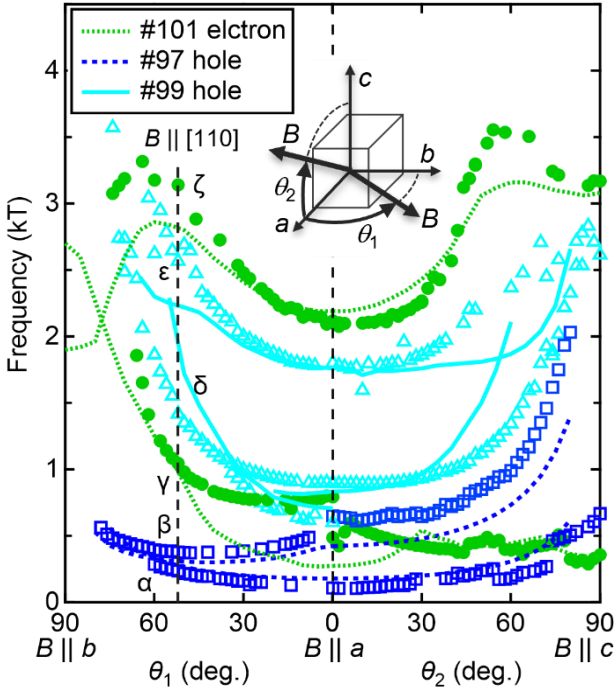

**Figure 2.** Angular dependences of dHvA frequencies with rotation axes along the *c* axis (the angle $\theta_1$) and *b* axis (the angle $\theta_2$). The angle of the applied magnetic field is defined as 0° at *B* ∥ *a*. Frequencies considered belonging to the same branch are plotted with the same symbol and color. The simulated angle dependence of the dHvA frequencies from the Fermi surface obtained through the first-principles calculations are displayed by the line with the same colour: the frequencies originating from #101 electron, #97 hole, and #99 hole are shown as dotted green, broken blue, and solid light blue lines, respectively. The inset image presents the schematic figure of the directions of magnetic fields.

## 3.3 Correspondence between the experiment and first-principles calculations

First-principles calculations were performed to reveal the correspondence between the FSs and angular dependence of dHvA oscillations. The crystallographic parameters of β-ReO$_2$ used in the first-principles calculations were precisely determined through synchrotron X-ray powder diffraction experiments (see appendix). Upon comparing the previously reported parameters[39] with the results of this study, we revealed that the atomic coordinates of the oxygen atoms differed by approximately 10%. The lattice constants differed from those optimised *via* first-principles calculations[37] by up to 1.6%.

The FSs obtained through first-principles calculations are consistent with previous reports[27], as illustrated in Fig. 3, and four FSs are present: the largest electron FS #101 electron centred at the *Γ* point, a small electron pocket #103 electron, and two hole FSs, #97 hole and #99 hole, covering the *R-T* line.

Figure 2 presents a comparison of the simulated dHvA oscillations from the calculated FSs with the experimental results. For the three relatively large FSs #101 electron, #97 hole, and #99 hole, both the experimental and simulated frequencies and the angular dependence are in good agreement. The agreement is particularly remarkable for *B* ∥ *a*, and the deviation increases as the angle moves away from *B* ∥ *a*. Considering the #101 electron, as shown in Fig. 3b, a simple orbit on the cylinder surrounding the *Γ* point labelled ζ gives extremal cross-sectional area at *B* ∥ *a*, where the calculation and experiment are in good agreement. However, as the magnetic field tilts away from *B* ∥ *a*, the shape of the orbit giving extremal cross-sectional area increases in complexity owing to the influence of the part extending to the *U* point (Fig. 3c), and a small difference in the FS shape causes a large difference in the value of the extremal cross-sectional area. Consequently, the discrepancy between theory and experiment becomes large in the region where $\theta_1$ and $\theta_2$ are large.

Based on the calculations, the dHvA oscillations originating from the smallest #103 electron are estimated to be less than 200 T. Previous magnetisation measurements have reported a QO with a frequency of 51 T at *B* ∥ *a*, which is believed to originate from the #103 electrons[27]. However, in the magnetic torque experiments in this study, the peaks could not be separated because of experimental noise superimposed onto the low-frequency region, and the angular dependence could not be clarified. Therefore, as described in section 3.5, we measured the Shubnikov-de Haas (SdH) oscillations at higher temperatures to investigate the shape and cyclotron mass of the #103 electrons.

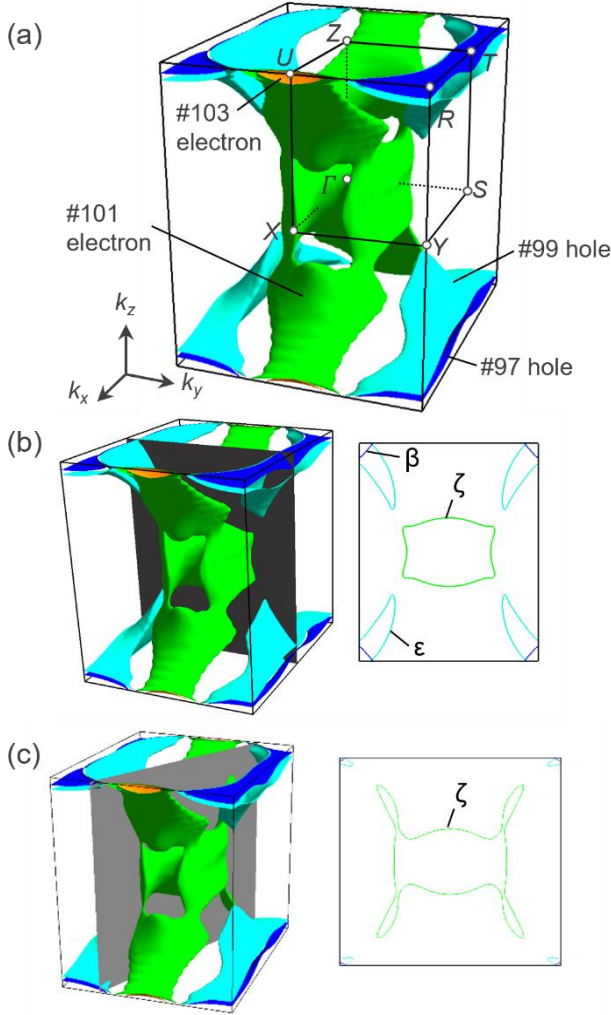

**Figure 3.** (a) Calculated Fermi surfaces of β-ReO$_2$ consisting of four Fermi surfaces (#97 hole, #99 hole, #101 electron, and #103 electron). Cross section of the Fermi surfaces crossing $\Gamma$ point perpendicular to the directions where (b) $\theta_1 = 0°$ and $\theta_2 = 0°$ (the [100] direction) and (c) $\theta_1 \sim 22°$ and $\theta_2 = 0°$. The labelles β, ε, ζ, in (b) and (c) are assigned in the FFT spectrum of the dHvA oscillations

## 3.4 Electron masses of Fermi surfaces

Figure 4 depicts the temperature dependence of the dHvA oscillation amplitude divided by the temperature for the representative branches at $B \parallel a$. The peaks at 2120, 900, and 597 T correspond to the calculated values of 2185 T (#101 electron), 833 T (#99 hole), and 422 T (#97 hole), respectively. The cyclotron mass ($m^*$) of the carrier was estimated by fitting the temperature dependence to the temperature decay factor $R_T$ using the Lifshitz-Kosevich (LK) equation.

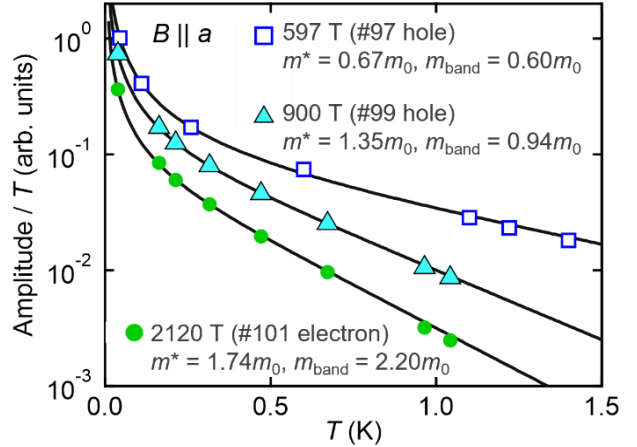

**Figure 4.** FFT amplitude of dHvA oscillations in the field range between 5 and 10 T divided by temperature ($T$) as a function of temperature. Solid black curves indicate the fitting to $1/\sinh(\kappa m^* T / B)$. The cyclotron masses estimated from the temperature dampings ($m^*$) and first-principles calculations ($m_{\text{band}}$) are shown for each branch in the unit of free electron mass ($m_0$).

The temperature dependences of the amplitudes are well fitted using the LK formula, in which the amplitude $Amp$ is proportional to the thermal damping factor ($R_T$) and Dingle damping factor ($R_D$), as $Amp \propto R_T R_D$; $R_T = (2\pi^2 k_B T m^* / eB\hbar)/\sinh(2\pi^2 k_B T m^*/eB\hbar)$ and $R_D = \exp(-2\pi^2 k_B T_D m^*/eB\hbar)$, where $k_B$ is the Boltzmann constant and $T_D$ is the Dingle temperature[40].

The obtained cyclotron masses are $1.74m_0$, $1.35m_0$, and $0.67m_0$ for the #101 electron, #99 hole, and #97 hole, respectively, where $m_0$ is the mass of free electron. The experimental values agree reasonably well with the values obtained from first-principles calculations: $2.20m_0$, $0.94m_0$, and $0.60m_0$. The calculated results reproduce the electronic structure of β-ReO$_2$ not only in the shape of the FSs but also in the band dispersion.

Under the same $B \parallel a$ conditions, the cyclotron mass of #103 electron was determined to be $0.23m_0$[27]. Essentially, the three large Fermi surfaces are as heavy as or twice as heavy as the free electrons, whereas the quasiparticles from the small electron pocket have cyclotron masses of less than 1/4 those of the free electrons.

## 3.5 SdH oscillations at high temperatures

According to the LK equation, the larger the cyclotron mass, the larger will be the temperature decay of the QO amplitude. Therefore, at high temperatures, QOs originating from FSs with large cyclotron masses disappear, and only QOs originating from quasiparticles with small cyclotron masses are observed. We noted the difference in cyclotron masses among the four FSs and observed only QOs originating from the #103 electrons by performing measurements above 1 K.

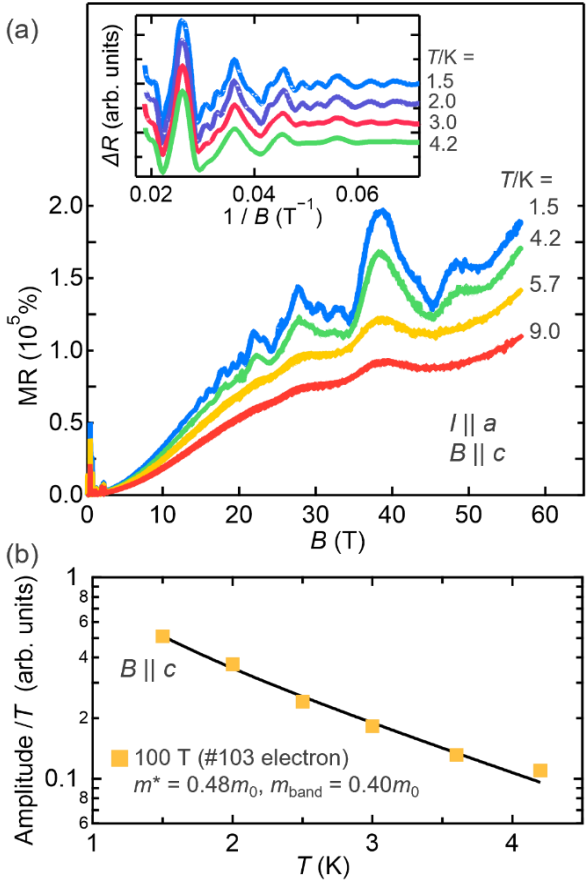

magnetic field, is the SdH oscillation. The SdH oscillations exhibit a complex magnetic-field dependence at 1.5 K, which is a combination of oscillations with large periods and fine oscillation components. In contrast, at 4.2 K, the fine component disappears and only a large period remains. This indicates that the QOs originating from the FSs with large cyclotron masses disappear and only the components originating from the FS with small cyclotron masses are observed.

After subtracting the monotonically increasing component of magnetoresistance using a polynomial function and performing an FFT, an oscillatory component at 100 T is observed up to temperatures above 4 K. The cyclotron mass obtained by fitting the temperature dependence of the amplitude from 1.5 to 4.2 K is $0.48m_0$ (Fig. 5b), which is in good agreement with the calculated value of $0.40m_0$.

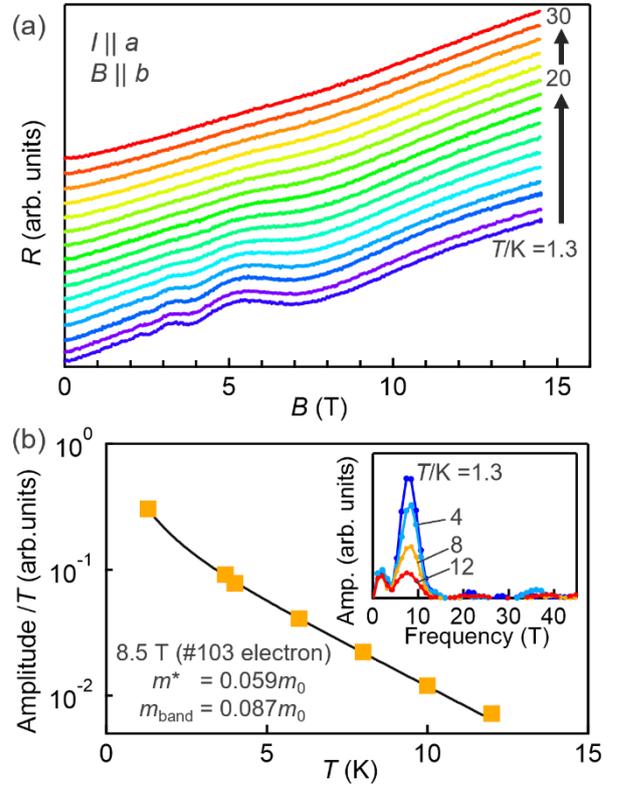

**Figure 5.** Magnetic-field dependence of transverse magnetoresistance (MR) measured with the electrical current $I$ flowing along the $a$ direction and magnetic field along the $c$ direction. The inset depicts oscillatory components after subtracting background plotted against the inverse of magnetic field. (b) FFT amplitude of SdH oscillations in the field range between 15 and 50 T divided by temperature ($T$) (orange square) as a function of temperature shown with a fit to $1/\sinh(\kappa m^* T / B)$ (black solid line). Cyclotron masses estimated from the temperature dampings ($m^*$) and first-principles calculations ($m_{band}$) are shown in the unit of free electron mass ($m_0$).

Figure 5a plots the transverse magnetoresistance data obtained by applying a current in the $a$ direction ($I \parallel a$) and a magnetic field in the $c$ direction ($B \parallel c$). At 1.5 K, the transverse magnetoresistance increased steeply with a magnetic field up to 30 T and slowly above 30 T up to 60 T, reaching a value 1,900 times higher at 60 T than under the absence of the magnetic field. The spike structures at low fields come from experimental noise. This field dependence is in contrast to that of WTe$_2$, a Weyl semimetal, which exhibits an increase in transverse magnetoresistance proportional to $B^2$ up to 60 T[41].

In the 1.5 K data, a distinct oscillation is superimposed on the magnetoresistance component above 10 T. This oscillation component, which increases in amplitude with an increasing

**Figure 6.** (a) Magnetic-field dependence of resistance measured with the electrical current $I$ flowing along the $a$ direction and magnetic field along the $b$ direction between 1.3 and 30 K. The curves are offset vertically for clarity. (b) FFT amplitude of SdH oscillations in the field range between 0.5 and 5.5 T divided by temperature ($T$) as a function of temperature. The solid black curve indicates the fitting to $1/\sinh(\kappa m^* T / B)$. The cyclotron masses estimated from the temperature dampings ($m^*$) and first-principles calculations ($m_{band}$) are shown for each branch in the unit of free electron mass ($m_0$). The inset depicts the fast Fourier transform of the oscillations of resistance in the range of magnetic fields of 1–14.5 T at various temperatures between 1.3 and 12 K.

The transverse magnetoresistance data for $I \parallel a$ and $B \parallel b$ is plotted in Fig. 6a. Clear SdH oscillations are observed at 1.5 K as in the $B \parallel c$ data. However, in this orientation, the QO is observed at a small magnetic field of 2 T. The oscillation period is also slow, suggesting that the QO originates from an extremely small FS. Oscillations are observed up to approximately 20 K, which is a high value for QO experiments.

The FFT of the oscillations yields a frequency of 8.5 T, and a cyclotron mass of $0.059m_0$ is obtained by fitting the amplitude to the temperature dependence. The FS corresponding to this frequency is considered to be the #103 electron, and the calculated frequency and cyclotron mass are 48 T and $0.087m_0$, respectively. The difference of approximately 5.6 times between the experimental and calculated frequencies may be attributed to the insufficient accuracy of the calculations for an extremely small FS.

## 4. Discussion

### 4.1 Relation between the small Fermi surface and the nodal chain

The dHvA and SdH oscillation results are in good agreement with the four calculated FSs: three large FSs #101 electron, #97 hole, and #99 hole with relatively heavy cyclotron masses, and a small electron pocket #103 electron with an extremely light cyclotron mass of at least $0.059m_0$.

The #103 electron is an electron pocket around the $U$ point in reciprocal space, which possesses a rugby-ball shape according to the calculations (Fig. 7a). Assuming #103 electron to be an ellipsoid, we estimated its actual shape using the cross sections obtained in the QO experiments: for $B \parallel a$, $b$, and $c$, the frequencies are 51, 8.5, and 100 T, respectively, and the extremal cross-sectional areas are $4.8 \times 10^{-3}$, $8.1 \times 10^{-4}$, and $9.6 \times 10^{-3}$ Å$^{-2}$, respectively. In particular, the cross-sectional area perpendicular to $k_y$ is only 0.13% of the area of the first Brillouin zone. The ellipsoid is shown schematically in Fig. 7a, where the ratio of the radius lengths in the $k_x$, $k_y$, and $k_z$ directions is 2:12:1, and the ellipsoid is long in the $k_y$ direction and collapses in the $k_z$ direction. In contrast, simulations based on first-principles calculations reveal that for $B \parallel a$, $b$, and $c$, the frequencies are 164, 48, and 187 T, respectively, and the ratio of the radii in the $k_x$, $k_y$, and $k_z$ directions is 1.1:4:1. Thus, although the length ratios differ, an elongated shape extending in the $k_y$ direction is consistent.

As shown in Fig. 7b, the #103 electron is formed through bands branching from the same point at the $U$ point, 85 meV below the $E_F$. Therefore, a small change in the $E_F$ causes a large change in the volume of the #103 electron. As the experimental extremal cross-sectional area is smaller than the calculated value, the actual $E_F$ is considered to be located between 0 and −85 meV of the calculation.

The reason for the anisotropic shape of #103 electron is that bands with remarkably different slopes extend from the same $U$ point in the $k_x$, $k_y$, and $k_z$ directions to form the FS. The slope of the band is reflected by the effective mass of the electrons. The cyclotron mass of electrons from the bands perpendicular to the $k_y$ direction, which have steep slope and small $k_F$, is very small ($0.059m_0$), whereas the cyclotron mass of electrons from the bands perpendicular to the $k_z$ direction, which have slow slope, is large ($0.48m_0$). The experimental and theoretical values are in good agreement, and their consistency further validates that the observed QOs originate from #103 electron.

The shape of the FSs near the $U$-$R$ line is closely related to the NC. β-ReO$_2$ has two types of nodal lines orthogonal in $k$-space (Fig. 7a): the first type is the nodal line 1 centred at the $U$ point on the $k_z = \pi$ plane (*ZURT* plane), and the second type is the nodal line 2 centred at the $R$ point on the $k_x = \pi$ plane (*UXSR* plane); the nodal lines 1 and 2 are protected by $n$- and $b$-glide symmetries, respectively. These two nodal lines intersect at a point $(\pi, 0.26\pi, \pi)$ on the $U$-$R$ line to form a NC extending in the $k_y$ direction. The results of calculations in this study are consistent with the initial theoretical work which predict NCs in β-ReO$_2$[25], although there are small quantitative differences between them: for example, the energies of neck-crossing Dirac points differ by several tens of meV. According to the present *ab initio* calculations, both nodal lines possess a considerably elongated shape: nodal line 1 crosses the $U$-$Z$ line at $k_x = 0.984\pi$, and nodal line 2 crosses the $R$-$S$ line at $k_z = 0.982\pi$.

In the band dispersion near the $E_F$, as shown in Fig. 7b, there exist hourglass-shaped band dispersions forming the nodal lines 1 and 2 on the $U$-$Z$ and $R$-$S$ lines, respectively, and their neck-crossing Dirac points lie at −105 meV and 136 meV. On the $U$-$R$ line, the intersection of nodal lines 1 and 2, where the two hourglass dispersions overlap, lie at −73 meV. As the actual $E_F$ is estimated to be between 0 and −85 meV, the NC exists at energies approximate to the $E_F$. The calculations considering SOIs in this study showed no gap opening at the Dirac point, thereby validating the protection bestowed by the glide symmetry.

Linear bands extending from the Dirac points originating from nodal lines 1 and 2 on the $U$-$Z$ and $R$-$S$ lines form the #101 electron and #99 hole (Fig. 7b). #101 electron and #99 hole are connected on the $U$-$R$ line in the cross section of the FSs at $k_z = \pi$ and $k_x = \pi$, forming a chain-like structure extending in the $k_y$ direction (Fig. 7c). In addition, #103 electron, which is surrounded by #101 electron in the $k_z = \pi$ plane, possesses a more elongated shape than that calculated, suggesting that nodal line 1 also has an even more elongated ring than that calculated.

If the NCs are present near the $E_F$, the Dirac electrons originating from the NCs must contribute to the transport properties. β-ReO$_2$ exhibits an extremely low electrical resistivity of 206 nΩcm at 2 K, which most likely results from the ultra-high mobility of Dirac electrons at low temperatures. In addition, β-ReO$_2$ is a metal with a large electron carrier density of $1 \times 10^{22}$ cm$^{-3}$, and such metals usually do not exhibit a large transverse magnetoresistance[27]. Although the transverse magnetoresistance of many topological semimetals is explained through electron-hole compensation[41], such an explanation is not applicable for β-ReO$_2$, where the electron and hole carrier densities are not compensated. The origin of the extremely large transverse magnetoresistance in β-ReO$_2$ may be that the topological protection acting on Dirac electrons is broken by the magnetic field[42].

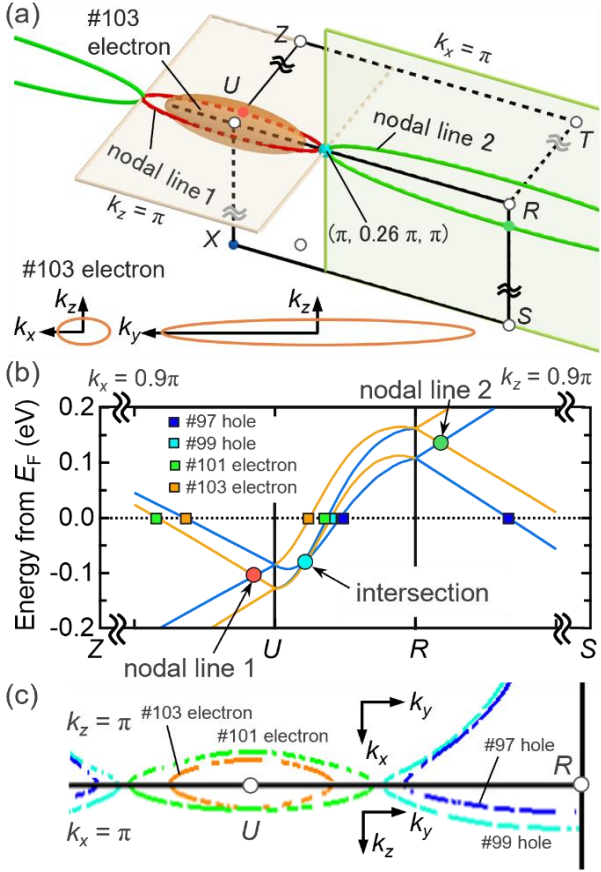

**Figure 7.** (a) Schematic figure of the geometric relation between the nodal chain and #103 electron pocket. At bottom left, cross-sections of #103 electron pocket determined from quantum oscillations assuming a flattened rugby ball shape is also shown (orange ellipses). (b) Electronic band dispersions around the $E_F$ along the $Z$–$U$, $U$–$R$, and $R$–$S$ lines showing hourglass dispersions with fourfold degenerate Dirac points at the neck crossings (red circle for nodal line 1, green circle for nodal line 2, and light blue circle for the intersection of nodal lines). The two pairs of bands in different colors (blue and orange) belong to different irreducible representations. The four Fermi surfaces and corresponding bands are indicated by different colored squares. (c) Cross section of the Fermi surfaces on the $k_z = \pi$ and $k_x = \pi$ planes near the $U$ and $R$ points.

The geometry of the NCs may produce a large azimuthal dependence on the transverse magnetoresistance. $B \parallel c$ and $a$, in which the magnetic fields are applied perpendicular to nodal lines 1 and 2, should have different effect on the Dirac electrons from $B \parallel b$, in which the magnetic field is applied parallel to both nodal lines. Detailed measurements of the magnetic-field-orientation dependence of the transverse magnetoresistance are expected to clarify the correlation between NC and transport properties. Moreover, the $E_F$ can be potentially tuned by controlling the oxygen content or chemical substitution to alter the contribution of nodal lines 1 and 2 or to maximise the transport properties derived from NCs.

### 4.2 Comparison with other topological materials

In topological semimetals, quasiparticles originating from Dirac or Weyl points have extremely light cyclotron masses. Table I compares the observed cyclotron masses with those of representative topological semimetals. The cyclotron mass of $0.059m_0$ observed in β-ReO$_2$ is comparable to those of other topological semimetals, suggesting the presence of Dirac electrons. Notably, #103 electron is not formed in a band branched from NC but is derived from a band crossing at the $U$ point, which is a time-reversal-invariant momentum. Therefore, the extremely small cyclotron mass of the #103 electron may reflect the nature of the Dirac electrons originating from band crossing at the $U$ point.

Table I. Comparison of transport properties among various types of topological semimetals (SMs).

| Material | Topological state | cyclotron mass ($m_0$) | MR(%) | Ref. |
|---|---|---|---|---|
| Cd$_3$As$_2$ | Dirac SM | 0.044 | $1.6 \times 10^3$ (1.5 K, 14.5 T) | 43) |
| NbP | Weyl SM | 0.076 | $8.5 \times 10^5$ (1.85 K, 9 T) | 44) |
| ZrSiS | Nodal line SM | 0.025-0.1 | $1.4 \times 10^5$ (2 K, 9 T) | 45),46) |
| β-ReO$_2$ | Nodal chain metal | 0.059 | $2.2 \times 10^4$ (2 K, 10 T) | This work |

Several candidate NC metals have been proposed; however, only a few candidates have been experimentally verified; TiB$_2$[47),48)] and Co$_2$MnGa[49)] have been verified as NC metals based on ARPES experiments. The NC in TiB$_2$ is protected by space- and time-reversal symmetry, which opens a gap of approximately 20 meV at the band crossing point when the SOIs are considered. In contrast, the NC of β-ReO$_2$ protected by the glide symmetry is stable to SOIs and does not open a gap. The difference between the two materials is expected to appear in the transport properties, which are significantly affected by the electronic state near $E_F$. A comparison of the two materials will yield a better understanding of the characteristics of NCs formed from different origins. Co$_2$MnGa is a ferromagnet with a $T_C$ of 690 K[50)], and its time-reversal symmetry is broken at room temperature. Thus, a comparative study of β-ReO$_2$ and Co$_2$MnGa is expected to clarify the difference between Dirac and Wyle NCs.

For both TiB$_2$ and Co$_2$MnGa, theoretically predicted drumhead surface states have been observed[47),49)]. In β-ReO$_2$, which is considered to be an NC metal, a drumhead surface state has been also predicted by theory[25]. Further ARPES study will verify this prediction[25)], and the transport properties, and optical responses derived from the drumhead surface state are expected to be elucidated in the future.

Recently, magnetic torque measurements revealed that the development of magnetic susceptibility at low temperatures is attributed to the quadrupolar fluctuations[51)]. In rhenium oxides, multipole ordering is induced by strong SOIs, in both the insulators and in metallic compounds[52)–55)]. Moreover, determining whether the electron correlation effects observed in $\beta$-ReO$_2$ originate from the bulk bands or from drumhead surface state with a high density of states is a key research interest.

## 5. Conclusion

In this study, quantum oscillation measurements were performed to determine the Fermi surface of $\beta$-ReO$_2$, an hourglass nodal-chain metal candidate. Clear quantum oscillations are observed in both the magnetic torque and resistivity measurements and show field-angle and temperature dependences that correspond well to the results of first-principles calculations using structural parameters determined through synchrotron XRD measurements. The temperature dependence reveals a significant difference in the cyclotron mass depending on the Fermi surface and orientation. The cyclotron mass of the large Fermi surface is approximately the same as that of the free electrons, whereas the small Fermi surface contained extremely light electrons with a minimum mass of 0.059 times that of the free electrons. According to the present calculation, the small electron pocket with light electrons is present near the nodal line, and thus the consistency between the calculation and experimental results suggests that the nodal chain exists at energies approximate to the Fermi energy.

## Acknowledgements

PXRD experiments were conducted at the BL5S2 of Aichi Synchrotron Radiation Center, Aichi Science and Technology Foundation, Aichi, Japan (Proposals No. 202201033). This work was partly supported by Japan Society for the Promotion of Science (JSPS) KAKENHI Grant Numbers JP20H01858, JP22H01173, JP22H01178, JP22K13996, JP23H04860, and JP22H04462 (Quantum Liquid Crystals).

## Appendix. Crystal strucrture analysis via synchroton powder X-ray diffraction

The structural parameters of $\beta$-ReO$_2$ were precisely determined for first-principles calculations through synchrotron powder X-ray diffraction experiments. Since most of the single crystal samples were twinned, powder samples were used for the measurements. Although structural data have been reported from laboratory X-ray diffraction experiments in the past[39)], precise measurement using high-intensity and high-energy synchrotron radiation is necessary to accurately determine the oxygen positions in $\beta$-ReO$_2$, which is composed of the heavy element Re and the light element oxygen.

As shown in Fig. A1, the simulated pattern precicely reproduces the XRD pattern, including the small peaks originating from oxygen, and the goodness-of-fit parameter $S$ is 1.07, indicating the validity of the obtained structural parameters. The obtained lattice constants are $a = 4.80691(9)$ Å, $b = 5.6400(1)$ Å, and $c = 4.59576(7)$ Å, which are close to previously reported experimental values $a = 4.812(1)$ Å, $b = 5.616(6)$ Å, and $c = 4.610(1)$ Å, with a difference of 0.1~0.4% (Table AI).

The theoretical study that pointed out the presence of NC have used a structural parameter optimized *via* first-principles calculations[37)], which differs from the experimentally obtained crystal structure data[39)]. The lattice constants obtained through first-principles calculations are large for all orientations, $a = 4.8861$ Å, $b = 5.7028$ Å, and $c = 4.6298$ Å, which differ by 0.7~1.6%.

Next, we compare the atomic positions of Re, for which the only variable is the $y$-coordinate, which in this study is 0.1082(2). The difference is 1.7 and 0.6%, respectively, which is closer to the first-principles calculation, with the previously reported value of 0.11 and the first-principles calculation of 0.1075. The atomic coordinates of oxygen in this study are 0.248, 0.365, and 0.416, which differ by 9.9% in the $z$-coordinate compared to the values of 0.25, 0.36, and 0.375 reported previously. Compared to the values 0.243, 0.360, and 0.4086 optimised *via* first-principles calculations, the difference is relatively small (1.4~2.0%).

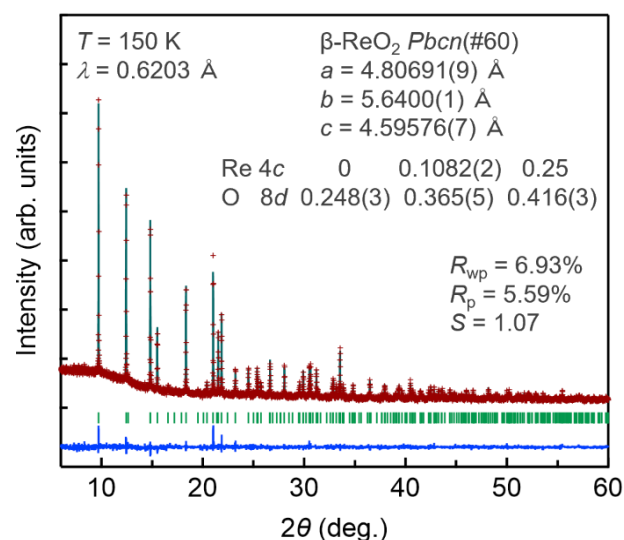

**Figure A1.** Synchrotron XRD pattern of a crushed single crystal sample of $\beta$-ReO$_2$ at 150 K and Rietveld fitting. Observed (red crosses), calculated (black solid line), and difference (lower blue solid line) XRD patterns are shown. Green tick marks indicate the position of allowed reflections.

Table AI. Comparison of crystallographic parameters between this study, previous repot[39] and those optimized *via* first-principles calculations[37].

|  | This study | Previous study[39] | First-principles calculation[37] |
|---|---|---|---|
| $a$ (Å) | 4.80691(9) | 4.812(1) | 4.8861 |
| $b$ (Å) | 5.6400(1) | 5.616(6) | 5.7028 |
| $c$ (Å) | 4.59576(7) | 4.610(1) | 4.6298 |
| $y$ (Re) | 0.1082(2) | 0.11 | 0.1075 |
| $x$ (O) | 0.248(3) | 0.25 | 0.243 |
| $y$ (O) | 0.365(5) | 0.36 | 0.360 |
| $z$ (O) | 0.416(3) | 0.375 | 0.4086 |